\documentclass[a4paper,twoside]{article}
\newcommand{\affil}[1]{$^{\rm #1}$}

\usepackage[authoryear]{natbib}
\bibpunct{(}{)}{;}{a}{}{,}

\usepackage{graphicx}

\date{} 

\def\spose#1{\hbox to 0pt{#1\hss}} 
\def\simlt{\mathrel{\spose{\lower 3pt\hbox{$\mathchar"218$}} 
\raise 2.0pt\hbox{$\mathchar"13C$}}} 
\def\simgt{\mathrel{\spose{\lower 3pt\hbox{$\mathchar"218$}} 
\raise 2.0pt\hbox{$\mathchar"13E$}}}

\title{\large\bf\flushleft The Lowest Metallicity Stars in the LMC:\\
Clues from MaGICC Simulations}

\author{\parbox{\textwidth}{\flushleft
\vspace{-0.5cm}
{\it Chris B. Brook,\affil{A} Maider S. Miranda,\affil{A,B} Brad. K. Gibson,\affil{B} Kate Pilkington\affil{B} and Greg S. Stinson\affil{C}}\\
\vspace{0.4cm}
{\small \affil{A}\,Departamento de F\'{i}sica Te\'{o}rica, Universidad Aut\'{o}noma de Madrid, E-28049, Madrid, Spain}\\
{\small \affil{B}\,Jeremiah Horrocks Institute, University of Central Lancashire, Preston, PR1~2HE, UK}\\
{\small \affil{C}\,Max-Planck-Institut f\"ur Astronomie, K\"onigstuhl 17, 69117, Heidelberg, Germany  }}}

\begin{document}

\twocolumn[
\begin{changemargin}{.8cm}{.5cm}
\begin{minipage}{.9\textwidth}
\vspace{-1cm}
\maketitle

\small{\bf Abstract:}
Using a cosmological hydrodynamical simulation of a galaxy of similar 
mass to the Large Magellanic Cloud (LMC), we examine the predicted 
characteristics of its lowest metallicity populations. In particular, we 
emphasise the spatial distributions of first (Pop~III) and second 
(polluted by only immediate Pop~III ancestors) generation stars. We find 
that primordial composition stars form not only in the central galaxy's 
progenitor, but also in locally collapsed sub-halos during the early 
phases of galaxy formation. The lowest metallicity stars in these 
sub-halos end up in a relatively extended distribution around the host, 
with these accreted stars possessing present-day galactocentric 
distances as great as $\sim$40~kpc. By contrast, the earliest stars 
formed within the central galaxy remain in the inner region, where the 
vast majority of star formation occurs, for the entirety of the 
simulation. Consequently, the fraction of stars that are from the 
earliest generation increases strongly with radius.


\medskip
\medskip
\end{minipage}
\end{changemargin}
]
\small

\section{Introduction} 
\label{ss.intro} 

The baryonic component of the Large Magellanic Cloud (LMC) is dominated 
by its large off-centre bar and underlying stellar disc 
($\sim$2.7$\times$10$^{9}$~M$_{\odot}$) \cite[][]{vandermarel02}, in 
addition to its associated neutral gas component 
($\sim$0.5$\times$10$^9$~M$_\odot$).  The bar is dominated by young 
stars \citep[][]{holtzman99,harris09} and shows a dearth of stars with 
ages of $\sim$5$-$12~Gyr. Photometric surveys of the disc also provide 
evidence for a comparably quiescent period of star formation at 
intermediate ages \citep[e.g.][]{smeckerhane02,harris09}.  Beyond the 
disc region, empirical evidence for the presence of an extended, and 
old, stellar halo (reaching to at least $\sim$25~kpc) also now exists 
\citep{munoz06}.

The distribution of the oldest and the most metal-poor stars in galaxies 
provides a unique ``near-field'' probe of the cold dark matter 
($\Lambda$CDM) hierarchical framework, with their relative numbers 
(compared with the bulk of the metallicity distribution function (MDF) 
providing insight into the complex interplay between accretion, 
outflows, star formation efficiencies, and internal mixing processes - 
e.g., \citealt{pilkington2012b}).  Indeed, the search for the most 
metal-poor stars in the Milky Way remains an important goal of 
contemporary astronomy \citep[e.g.][and references therein]{norris2013}.  
Analogous searches for such first or second generation stars within the 
LMC are obviously hampered by the latter's distance ($\sim$50~kpc: 
\citealt{Gibson2000}).  Having said that, neither the stellar component 
of the bar region (traced by AGB stars: \citealt{cole05}) nor that of 
the older underlying population (traced by RR~Lyrae: 
\citealt{haschke12}) have thus far shown evidence for the presence of 
stars with metallicities below [Fe/H]$\sim$$-$3.\footnote{Although one 
might not expect to find RR~Lyrae stars with [Fe/H]$<$$-$3, as such 
extremely metal-poor stars may not enter the instability strip 
\citep{yoonlee02}.}

Beyond the extreme metal-poor tail of stellar abundances within the LMC, 
there also exists empirical evidence of both an age-metallicity 
\citep{holtzman99,cole05,rubele11} and age-radius \citep{piatti13} 
gradient, with essentially no significant radial metallicity gradient 
\citep{cioni09,cole09,feast10}. In consort, such trends support a 
scenario in which the LMC formed in an ``outside-in'' manner (cf. the 
canonical ``inside-out'' scenario usually associated with the Milky Way 
- e.g., \citealt{pilkington2012a}).

Significant progress has been made in relation to predicting the 
expected distribution of old and metal-poor stars within our Galaxy, 
using Milky Way-scale disc simulations.  Both \citet{white00} and 
\citet{diemand05} suggest that the central regions today are the best 
place to search for the \it oldest \rm stars.  However, the link between 
the oldest stars, metal-free (Pop~III) stars, and extremely metal-poor 
(pseudo-second generation) stars is not so straightforward.

In \citet{scannapieco06} and \citet{brook07}, we demonstrated that 
Pop~III stars formed over a range that peaked at $z$$\sim$10, but 
continued to form down to $z$$\sim$5, and were distributed over a wide 
range of galactocentric radii at $z=0$. Under the assumption that the 
Pop~III initial mass function (IMF) allowed stars with appropriately low 
mass to survive until the present-day, our prediction was that they 
would be distributed throughout the Galactic halo.  On the other hand, 
the oldest stars formed in halos that collapsed close to the highest 
density peak of the final system, and at $z$=0 were located in the 
central/bulge region of the Galaxy. These two studies also predicted 
similar distributions for second generation stars - i.e., those enriched 
only by Pop~III stars - and showed that the {\it fraction} of first and 
second generation stars compared to the total stellar population was 
{\it higher} in the extended halo, compared with the bulge.

The question now arises; will old and metal poor stars be distributed in 
a similar manner in lower mass galaxies, such as the LMC? Hierarchical 
build up of mass is (almost) self-similar with respect to different mass 
halos in a $\Lambda$CDM Universe \citep[e.g.][]{maccio08}. That said, 
the stellar mass-halo mass relation \citep{moster10,guo10} and the 
missing satellite problem \citep{klypin99,moore99} suggest that baryon 
fraction is a strong function of total mass, and that there may be a 
mass below which dark matter halos have no baryons.\footnote{The 
existence of baryons in low mass halos may be a combination of collapse 
time and mass, rather than mass alone, with halos that collapse later 
less likely to house baryons \cite[see e.g.][]{bullock00}.} Should the 
LMC have an accreted population of stars, given its mass? According to 
\citet{purcell07}, the answer would be yes, with an accreted diffuse 
stellar component of mass $\sim$10$^7$~M$_\odot$. Will the accretion 
events responsible for this component also bring in first/second 
generation stars and, if so, how will they be distributed in an 
LMC-like system?  In this study, we examine a fully cosmological 
simulation of a dwarf irregular which has a comparable stellar mass to 
that of the LMC, in order to answer these questions.

A brief overview of the code and the underlying physics associated with 
the simulation employed here is provided in \S\ref{sims}. We then (\S\ref{res}) examine the resolution of the simulations and  relation between the sites of first star formation in our simulations to what is found in specialised high resolution simulations focussing on early star formation. The 
primary results (\S\ref{results}) and conclusions (\S\ref{conclusions})  are
drawn thereafter.

\section{The Simulation}
\label{sims}

We make use of initial conditions drawn from the MacMaster Unbiased 
Galaxy Simulations (MUGS: \citealt{stinson10}), but employ the feedback 
prescription outlined by the Making Galaxies in a Cosmological Context 
collaboration (MaGICC: \citealt{brook12b}). We inject 10$^{51}$ erg per 
supernova in the form of thermal energy into the surrounding 
interstellar medium (ISM), using the blastwave formalism of 
\citet{stinson06}.\footnote{The radius over which the energy is 
deposited corresponds to radius at which the interior pressure in the 
remnant is reduced to that of the pressure of the ambient ISM - i.e., 
R$_{\rm merge}$, using the terminology of \citet{Gibson1994}.}  Cooling 
is disabled for gas particles situated within a blast region of size 
$\sim$100~pc, for a time period of order 10~Myr.\footnote{This is the 
timescale for the remnant's interior to cool to $\sim$10$^4$~K, below 
which radiative losses are minimal; this corresponds roughly to a 
timescale $\sim$200$\times$ that associated with the formation of the 
thin dense shell, subsequent to the radiative cooling of the shocked 
material (the latter occurs near the end of the Sedov-Taylor phase) - 
\citep{Gibson1994}.} We adopt a \citet{chabrier03} IMF, and include the 
effects of radiation energy from massive stars in the $\sim$4~Myr prior 
to the appearance of a stellar particle's first Type~II supernova. The 
efficiency with which this latter energy couples to the ISM, though, is 
$<$1\%.  Star formation is restricted to regions which are both 
sufficiently cool ($<$15000K) and dense ($>9.3$~cm$^{-3}$).

The initial gas (dark matter) particle mass of the simulation is 2.4$\times$10$^4$M$_\odot$ (1.7$\times$10$^5$M$_\odot$), while stars are  formed with mass 7.9$\times$10$^3$M$_\odot$. 
The gravitational softening length is 155\,pc for all particles.

The  simulation employed here - our LMC analogue - was 
utilised in our earlier work to study its global photometric and 
kinematic characteristics\footnote{Where it was referred to by the label 
SG3.} \citep{brook12b}, the chemical properties of its stellar disc 
\citep{brook12b}, and the skewness, kurtosis, and higher-order moments 
of its MDF\footnote{Where it was referred to by the label 11mChab.} 
\citep{pilkington2012b}.  Its stellar mass 
(4.2$\times$10$^{9}$~M$_\odot$) and rotation velocity ($\sim$85~km/s at 
2.2 disc scalelengths) means it is somewhat more massive than the LMC 
and, as such, all comparisons are made with this caveat in mind.

The simulation, along with its companions within the MaGICC 
suite, are consistent with the empirical stellar mass-halo mass relation 
\citep{brook12b,stinson13} as derived by halo-matching studies 
\citep{moster10,guo10}, as well as the evolution of such relations 
\citep{moster13,stinson13}; in addition, they match a wide range of 
galaxy scaling relations \citep{brook12b}. This gives a degree of 
confidence in the history of the build up of the baryonic mass within 
the dark matter dominated growth of the galaxy mass, and in particular 
that there are reasonable constraints on the accretion of stellar 
systems when mergers occur. Further, our simulations include strong 
outflows \citep{brook11}, particularly in dwarfs, and reproduce the extended 
metal enrichment of the circumgalactic media of intermediate mass 
galaxies \citep{stinson12}, and possess temporally invariant metallicity 
gradients \citep{pilkington2012b,gibson13}.

\begin{figure}
\includegraphics[height=.25\textheight]{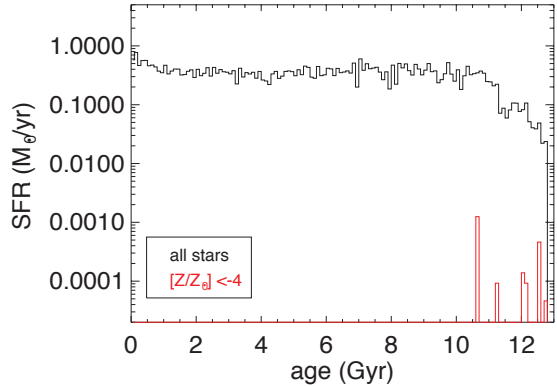}
\caption{Star formation history of our LMC analogue, using all stars 
within the virial radius, colour-coded in black (for all stars) and red 
(extremely metal-poor stars: i.e., Z$<$Z$_\odot$/10,000). Here, the 
extremely metal-poor stars form in early, discrete, bursts, while the 
star formation history for the ensemble is fairly constant over the past 
$\sim$10~Gyr.}
\label{sfr}
\end{figure}

Extended stellar halos in low mass galaxies may 
form from processes other than accretion. \citet{stinson09} show that 
elevated early star formation activity combined with supernova feedback 
can produce an extended stellar distribution in the absence of 
accretion, while \citet{zolotov09} and \citet{house11} show that stars 
formed \it in situ \rm can also contribute to stellar halos. These 
processes  are self-consistently 
included in our simulations, and are thus accounted for in our analysis 
of the distribution of low metallicity populations. Further, in this 
study, we analyse an isolated galaxy formation simulation that is the 
closest analogue to the LMC in terms of stellar mass of those formed in 
the MaGICC program. Thus, an assumption of this paper is that the early 
stars in the LMC are not enriched by outflows from the progenitor of the 
Milky Way. Given the analysis of \cite{scannapieco06}, who showed that 
even in models with extreme outflows, lower mass objects which collapse 
away from the central galaxy are formed from primordial gas, this would 
appear to be a reasonable assumption, particularly if the LMC is truly 
on its first orbit around the Milky Way \citep{Besla07}.

\begin{figure}
\includegraphics[height=.33\textheight]{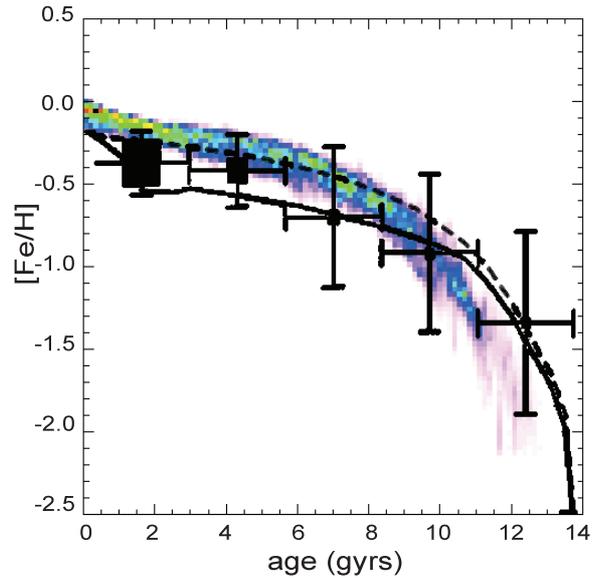}
\caption{Age-metallicity relation for all stars within the virial radius 
of our simulated LMC analogue. The colour-coding of the contours is by 
density, ranging from purple (fewest stars) to green (most stars). The 
black curves and points represent analytical models and empirical data 
for the LMC, after \citet{pagel98}.}
\label{age}
\end{figure}

Finally, we note that our simulation does not incorporate an event which 
results in an analogue of the significant bar seen in the LMC. Instead, 
our study aims to explore the formation and evolution of the older 
populations and underlying disc stars. While not meaning to be a 
one-to-one `clone' of the LMC, we will see that our model can provide 
some broad insights into the lowest metallicity populations of stars 
expected of galaxies of this mass.

\section{Earliest star forming sites}
\label{res}
The nature of the sites of first star formation remains uncertain and the subject of significant research. Early analytic models found variously that the earliest stars form within halos of mass between $\sim 5$$\times$10$^{4}$M$_\odot$, and  $\sim 5$$\times$10$^{7}$M$_\odot$ at  redshifts between 100 and 10 \citep{tegmark97}. Isolated simulations favoured halo masses of  $\sim 7$$\times$10$^{5}$M$_\odot$  \citep{fuller00}, while cosmological simulations favoured   $\sim 7$$\times$10$^{5}$M$_\odot$ \citep{yoshida03} and found no dependence on redshift.  

Recent simulations continue to apply increasingly sophisticated models, including detailed radiative feedback, with \cite{umemura13} finding that multiple stars may form within halos of mass $\sim$10$^{5}$M$_\odot$. \cite{xu13} study Pop III star formation at high resolution over a relatively large volume, and find that  Pop III stars form in halos between 4$\times$10$^{6}$M$_\odot$ and 3$\times$10$^{8}$M$_\odot$, with the majority forming in halos with masses of a few $\times$10$^{7}$M$_\odot$. In their simulations, Pop III stars inside more massive halos are the result of mergers of small halos, rather than forming in the more massive  halos themselves. We note that significantly more work has been done on the earliest stars that form in the largest over-densities at high redshift, that will be progenitors of massive galaxies, than in  smaller over-densities that are progenitors of lower mass galaxies such as the LMC.

In our simulations, the first stars form within a halo with virial mass $\sim 5$$\times$10$^{7}$M$_\odot$, at $z$$\sim$13. Only a single star particle (7$\times$10$^{3}$M$_{\odot}$) that forms in this earliest collapsing halo has zero metallicity, and it is able to pollute the surrounding region such that subsequent star particles have some metals.  Just to make it clear,  our model is not capturing details of the progression of star formation and self- enrichment within these first star forming cites, which remains uncertain. Rather, our labels of zero metallicity stars  reflect the properties of the most metal-poor stars formed in any given region. Thus, it is entirely possible that some fraction of the star particle forming in halos that collapse from primordial gas will nevertheless contain metals that were inherited from other stars forming in the same temporally extended burst. For our purposes here then, stars formed in such self-enriched primordial star clusters are included in the distribution that we have labeled as zero metallicity, or ÔÔprimordial stars.ÕÕ  Similarly, if the first stars in LMC mass galaxies form in lower mass systems at higher redshift than what we resolve, then presumably these would be incorporated into these collapsing regions that we do resolve.  Yet the evolutionary fate of these stars, whether they are primordial or simply low metallicity, is well modelled in our simulations.   

The formation of primordial stars in our simulation then proceeds as independent proto-galactic regions collapse to $5-8$$\times$10$^7$M$_\odot$. We have identified that 14 such regions that house primordial star formation exist , and  that they collapse between redshift $\sim$13 and redshift  $\sim$3. The number of star particles in each independent cite of primordial star formation ranges from 1-3, meaning a mass of  (0.7-2.1$\times$10$^{4}$M$_{\odot}$). Again, some uncertainty remains as to how star formation should proceed in each of these sites. In particular, our simulations show no dependence of the mass of stars formed in these sites with redshift. Thus, {\it a caveat of our model is that the early self-pollution of the separate sites of  primordial star formation  will be relatively similar} and in particular are independent of redshift. A strong redshift dependence may have an effect on the shape of the radial distribution of early generation stars, as shown in section \S\ref{results}.

\section{Results}
\label{results}

\begin{figure}
\includegraphics[height=.25\textheight]{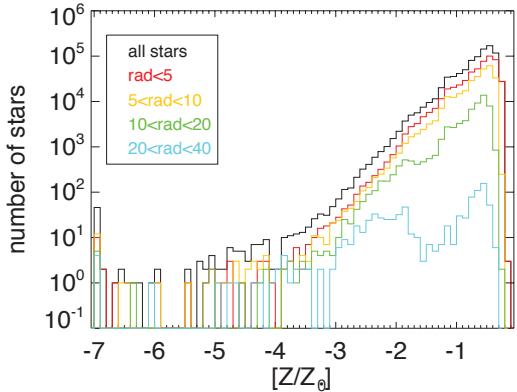}
\caption{Metallicity distribution function of disc stars within our 
simulated LMC analogue, as a function of galactocentric radius - black 
(all stars), red (radii $<$5~kpc), yellow (5$<$radii$<$10~kpc), green 
(10$<$radii$<$20~kpc), and cyan (20$<$radii$<$40~kpc).}
\label{mdf}
\end{figure}

In Figure~\ref{sfr}, we show the star formation history (SFH) for both 
the ensemble of stars within the virial radius (black) and the extremely 
metal-poor (EMP) population ([Z/Z$_{\odot}$]$<-4$, in red).  EMP stars 
are all older than $\sim$10~Gyr and form in 4$-$5 discrete `bursts', 
while the overall star formation rate is fairly constant over final 
$\sim$10~Gyr of the simulation (with a slight upturn in the final 
$\sim$300~Myr).\footnote{Fig\,\ref{sfr} is somewhat analogous to Fig\,1 of 
\citet{pilkington2012b}, although here, we sub-divide the ensemble and 
EMP populations, while in the latter, we only showed the star formation 
history of the analogous solar neighbourhood of 11mChab (blue curve, 
therein).}

In Figure~\ref{age}, we show the age-metallicity relation (AMR) of the 
ensemble population of stars in our LMC analogue.\footnote{Fig\,\ref{age} 
is analogous to the upper right panel of Pilkington et~al.'s (2012b) 
Fig\,2, albeit for the ensemble of stars rather than just the sub-set of 
`solar neighbourhood' stars shown in the latter.} The overall trend 
(both qualitatively and quantitatively) is remarkably similar to that 
seen for the LMC, in nature (overlaid in Fig\,\ref{age} with black 
symbols).  The curves in Fig\,\ref{age} correspond to the analytical 
models of \citet{pagel98}.

We next show the metallicity distribution functions (MDFs) as a function 
of galactocentric radius, for the ensemble of stars in our LMC analogue 
(the black line within Fig\,\ref{mdf}).\footnote{The black curve of 
Fig\,\ref{mdf} is analogous to the upper right panel of Pilkington 
et~al.'s (2012b) Fig\,3, albeit for the ensemble of stars rather than 
just the sub-set of `solar neighbourhood' stars shown in the latter.} We 
note firstly that the stellar halo of the galaxy extends out to 
$\sim$40~kpc.  The red line (radii $<$5~kpc) shows that a low 
metallicity tail exists in the main body of the galaxy, but that very 
few stars - \it fractionally speaking \rm - have [Z/Z$_{\odot}$]$<$-4 
(i.e., few EMPs). Conversely, as we move to larger radii, the \it 
fraction \rm of EMPs increases (even if the overall numbers do not).

\begin{figure}[t]
\includegraphics[height=.25\textheight]{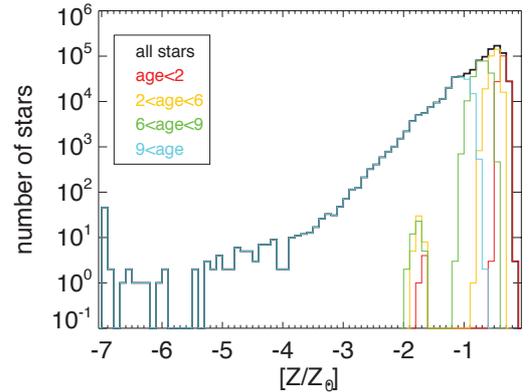}
\caption{Metallicity distribution functions for the ensemble of stars 
within the virial radius of our simulated LMC analogue, sub-divided by 
stellar age.  In black, all stars are shown; in red, only stars formed 
in the past 2~Gyrs are shown; yellow, green, and cyan correspond, 
respectively, to ages 2$\rightarrow$5~Gyr, 6$\rightarrow$9~Gyr, and 
$>$9~Gyr.}
\label{mdfage}
\end{figure}

In Fig\,\ref{mdfage}, the MDF is separated by stellar ages, with red 
being the youngest and cyan the oldest. Stars with [Z/Z$_{\odot}$]$<$$-2$ 
are older than 9\,Gyr. The metallicity decreases with increasing age, as 
per the age-metallicity relation (Fig\,\ref{age}). The very youngest 
stars (red line, age$<$2\,Gyrs) have a very narrow distribution at 
[Z/Z$_{\odot}$]$\sim$-0.5 and a few stars with [Z/Z$_{\odot}$]$\sim$-2. 
The oldest stars (light blue, ages$>$9\,Gyrs) have a peak at 
[Z/Z$_{\odot}$]$\sim$-1.0, with a long tail.  These trends are 
consistent with our earlier analysis of the `bulge' and `solar 
neighbourhood' regions of a comparable simulated dwarf (Figs\,4\,\&\,5 of 
Pilkington et~al. 2012b are analogous to that of Fig\,\ref{mdfage} here, 
albeit the latter now shows the full ensemble of stars rather than just 
 two spatially-selected sub-sets employed in Pilkington et~al.).

\begin{figure}
\includegraphics[height=.25\textheight]{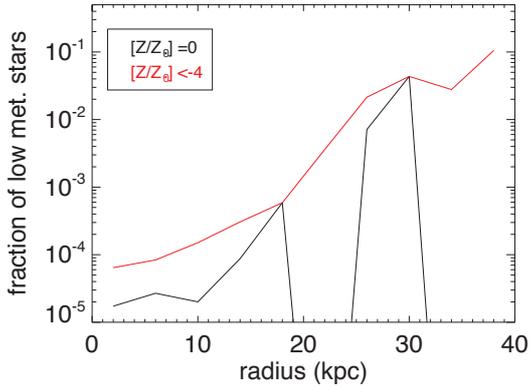}
\caption{Fraction of zero metallicity (black) and EMP (red) stars 
(normalised by the total number of stars at each galactocentric 
radius).}
\label{lowmet}
\end{figure}

In Fig\,\ref{lowmet}, we focus on the the zero 
metallicity (black) and EMP stars (red).  We show the fraction of 
low metallicity stars as a function of galactocentric radius (normalised 
by the total number of stars in each radial bin).  While the number 
statistics are low, as a fraction of the stellar population, low 
metallicity stars only become significant in the outer parts of the 
galaxy. We restate  the caveat mentioned in \S\ref{res} that the enrichment of the independent regions of primordial star formation are relatively uniform in our models, without a significant redshift dependence. Under this assumption,  the predicted trends in Fig \ref{lowmet}  will remain, although the absolute values of the fractions will vary according to the details of the earliest star formation within the earliest collapsing proto-galactic halos.

\begin{figure*}
\hspace{-1.cm}  
\includegraphics[height=.265\textheight]{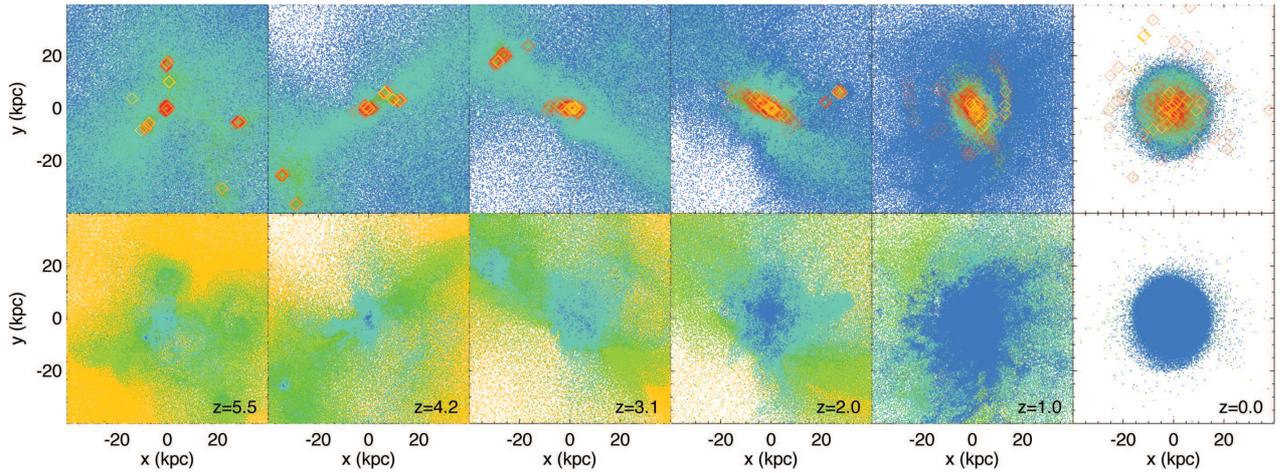}
\caption{Temporal evolution of all baryons which will form stars by 
redshift $z=0$ in the $x$-$y$ plane with sides of 40\,kpc.  The columns, 
from left to right, correspond to redshifts $z$= 5.5, 4.2, 3.1, 2.0, 
1.0, and 0, respectively.  Baryons in the top row are coloured by the 
metallicity of the stars which they will form by $z=0$. Baryons in 
the bottom row are coloured according to the metallicity at each time 
step. The colour distribution is the same in both sequences: 
[Z/Z$_\odot$]$>-1$ (blue), $-3<$[Z/Z$_\odot$]$<-1$ (cyan), 
[Z/Z$_\odot$]$<-3$ (green), and zero metallicity (yellow).  In the top 
row, we have identified separately the baryons (i.e. stars and the gas from which stars will form)  with 
[Z/Z$_\odot$]$<-4$ (yellow diamonds) and zero metallicty stars (red 
diamonds).}
\label{now}
\end{figure*}

We can now visualise the evolution of the galaxy and its metals, and 
identify the specific origin of the low metallicity stars. In 
Fig\,\ref{now}, time evolves from left to right, with the six columns 
corresponding to redshifts $z$ = 5.5, 4.2, 3.1, 2.0, 1.0 and 0, 
respectively. We plot all baryons that form stars by $z=0$; each point 
can be a gas element, or it may have already formed a star. Thus,  in the right-most column, only stars are shown because, by 
construction, it is redshift $z=0$; conversely, in the earlier time 
steps gas is also shown (which will later form stars), with the gas 
dominating increasingly  to the left (i.e., at  higher 
redshifts).

The color distribution is the same for both rows in Fig\,\ref{now}: the 
highest metallicity is represented in blue, and as metallicity decreases 
the color changes progressively to cyan, green, and, yellow. In the top 
row, each baryon is coloured according to this mapping (i.e., to the 
metallicity of the star it will have formed by $z=0$). Thus, in the top 
left panel, baryons that form the lowest metallicity stars (green, 
[Z/Z$_{\odot}$]$<$$-$3) are already collapsing in the densest regions. 
By contrast, the gas which will form stars with high metallicity (blue, 
[Z/Z$_{\odot}$]$>$$-$1) are in the outer, less dense regions at $z$=5.5, 
and will not form stars until later, by which time they are enriched.

In the bottom row of Fig\,\ref{now}, colours are assigned to the baryons 
by the metallicity they have at the timestep of each panel.  This shows 
the enrichment is most evolved in the central region, where stars are 
forming, with the $z$=4.2 panel, in particular, showing that enrichment 
is quite inhomogeneous, with enrichment proceeding around locally 
collapsed regions. The region around the central main progenitor is the 
most enriched. Note that the order of overplotting the different 
metallicities is opposite in each row: in the top panels the low 
metallicity baryons are overplotted on the ones with higher metallicity; 
in the bottom row we follow the inverse process. In other words, the two 
panels of the far-right column are the same, but plotted in the inverse 
metallicity order.

The yellow diamonds in both panels of Fig\,\ref{now} indicate 
baryons with zero metallicity  (stars or the gas from 
which they form). Red diamonds are the extremely metal-poor ([Z/Z$_{\odot}$]$<$$-$4) stars. Clearly, primordial gas is collapsing in several local 
regions and forming stars, prior to accretion to the central region. 
This process is what results in the more extended distribution of the 
lowest metallicity stars, compared to the general population, which is 
dominated by stars that form at later times from gas that accretes to 
the central region before forming stars.

\section{Conclusions}
\label{conclusions}

Cold dark matter cosmology is characterised by the hierarchical build-up 
of mass through merging processes. So long as stars are able to form 
within locally collapsing regions during an early rapid collapse phase, 
they will subsequently be thrown into an extended distribution (a 
stellar halo) as the central galaxy is assembled.  Using fully 
cosmological hydrodynamical zoom simulations of a dwarf galaxy, we have 
shown that we can expect this to occur in galaxies in the mass range of 
the LMC. In our simulation, the accreted stars extend out to 
$\sim$40~kpc in a system which has double the stellar mass of the LMC.

We have further shown that we can expect that the first stars forming 
within high-redshift, locally collapsing, regions, will form from 
primordial material, meaning that the first and second generation stars 
will have an extended distribution at $z=0$.  
This is in spite of the 
significant large scale supernova-driven outflows that occur in the 
simulated galaxies of our MaGICC program 
\citep{brook12b,brook12c,stinson12}. 
In our simulations, 14 independent sites of primordial star formation exist, as proto-halos collapse to $\sim 5-8$$\times$10$^7$M$_\odot$ between redshift $\sim$13-3. 
Later star formation is dominated by 
{\it in situ} star formation in the inner region of the galaxy, meaning 
that it is in the outer regions where the largest {\it fraction} of 
first and second generation stars will be found at $z=0$. So long as the enrichment of the independent regions of primordial star formation are relatively uniform, without a significant redshift dependence, these predicted  trends will remain, although the absolute values of the fractions will vary according to the details of the earliest star formation within the earliest collapsing proto-galactic halos. 
 
Although this comparison of a simulated galaxy using hydrodynamics 
within a cosmological framework has provided interesting insights into 
the expectations regarding the oldest populations in low mass galaxies 
such as the LMC, a larger statistical suite of simulations are required 
to make a firmer one-to-one comparison with the LMC-Milky Way system as 
a whole. We note that the extended population of stars in the LMC, as predicted in our simulations, evokes the predictions made by \cite{bovill11} of `ghost halos' around isolated dwarfs as remnants of primordial star formation.

\section*{Acknowledgments} 
CBB is supported by the MICINN (Spain) through the grant AYA2009-12792.
BKG acknowledges the support of the UKâ Science \& Technology 
Facilities Council (ST/J001341/1). The generous allocation of resources 
from STFC's DiRAC Facility (COSMOS: Galactic Archaeology) is gratefully 
acknowledged. We also thank the PRACE-2IP project (FP7 RI-283493) for 
support, in addition to the University of Central Lancashire's High 
Performance Computing Facility.

\bibliographystyle{apj}

\bibliography{LMC}

\end{document}